\documentclass[runningheads]{llncs}
\usepackage[T1]{fontenc}
\usepackage{graphicx}
\usepackage{amsmath}
\usepackage{amsfonts}
\usepackage{booktabs} 
\begin{document}
\title{An Emotion Recognition Framework via Cross-modal Alignment of EEG and Eye Movement Data}
%
%
\author{Jianlu Wang\inst{1}\orcidID{0009-0002-3757-2283} \and
Yanan Wang\inst{2}\orcidID{0000-0003-1919-686X} \and Tong Liu \inst{1}\orcidID{0000-0003-4947-1340}\thanks{Corresponding author: liu\_tongtong@foxmail.com}}
\authorrunning{Jianlu Wang et al.}
%
\institute{College of Computer Science and Engineering, Shandong University of Science and Technology, Qingdao 266590, China.
\email{202211070822@sdust.edu.cn,liu\_tongtong@foxmail.com}\\
\and
Graduate School of Computer Science and Engineering, University of Aizu, Aizuwakamatsu, Fukushima, 965-8580 Japan\\
\email{wang\_yanan99@outlook.com}}

%
\maketitle              
\vspace{-15pt}

\begin{abstract}
Emotion recognition is essential for applications in affective computing and behavioral prediction, but conventional systems relying on single-modality data often fail to capture the complexity of affective states. 
To address this limitation, we propose an emotion recognition framework that achieves accurate multimodal alignment of Electroencephalogram (EEG) and eye movement data through a hybrid architecture based on cross-modal attention mechanism. 
Experiments on the SEED-IV dataset demonstrate that our method achieve 90.62\% accuracy.
This work provides a promising foundation for leveraging multimodal data in emotion recognition.

\end{abstract}
\vspace{-27pt}
\section{Introduction}
\vspace{-5pt}
Human emotions play a crucial role in cognition and behavior, guiding decision-making and shaping social interactions. 
Automated emotion recognition, the computational identification of affective states from physiological or behavioral signals, has therefore attracted considerable attention in fields ranging from affective computing and healthcare analysis to driver monitoring and adaptive learning systems \cite{25calvo2010affect}. 
By accurately decoding a user’s emotional state in real time, emotion recognition systems can enable personalized feedback, enhance user engagement, and predict subsequent behaviors \cite{26hudlicka2003feel}.

Most existing emotion recognition systems rely solely on unimodal cues such as speech \cite{27koolagudi2012emotion} or facial expressions \cite{28tarnowski2017emotion}, which may be insufficient under challenging conditions. 
Among physiological modalities, EEG and eye movement data offer distinct advantages. 
EEG captures high-temporal-resolution cortical dynamics associated with emotional valence and arousal, while eye movements reflect attentional shifts and autonomic nervous system activity \cite{29zheng2015investigating}. 
Unlike speech or facial expressions, these modalities are less susceptible to social masking and provide complementary perspectives on emotional states. 
Nevertheless, integrating these modalities remains challenging due to noise, high dimensionality, and temporal misalignment \cite{1saxena2020emotion}.

To address these challenges, we propose an emotion recognition framework comprising three core components: (1) a hybrid feature selection module (PCA-RFE) for noise reduction and discriminative feature extraction, (2) a temporally-aware cross-modal attention mechanism to dynamically align EEG and eye movement sequences, and (3) a residual multi-layer perceptron (MLP) classifier for hierarchical emotion pattern refinement.

The remainder of this paper is organized as follows. 
Section 2 reviews related work on emotion recognition and multimodal learning. 
Section 3 introduces the details of the proposed method.
Section 4 presents experimental results and analysis.
Section 5 concludes the paper and outlines future research directions.

\vspace{-12pt}
\section{Related Work}
\vspace{-5pt}
\subsection{Emotion Recognition}
\vspace{-5pt}
Emotion recognition aims to identify human emotional states by analyzing behavioral and physiological signals, such as speech, facial expressions, or neurophysiological data \cite{1saxena2020emotion}. 
Traditional methods primarily relied on handcrafted feature engineering combined with Machine Learning (ML) methods, including Support Vector Machines (SVM) and Random Forests (RF) \cite{1saxena2020emotion}. 
However, with the advancement of Deep Learning (DL), models such as Convolutional Neural Networks (CNNs), Recurrent Neural Networks (RNNs), and Multilayer Perceptrons (MLPs) have demonstrated superior performance by automatically learning hierarchical feature representations from raw data \cite{7zhang2024deep}.
These Deep Learning-based approaches have become fundamental in advancing emotion recognition task by effectively capturing complex patterns in behavioral and physiological data \cite{6jafari2023emotion}.

\vspace{-10pt}
\subsection{Multimodal Learning}

Multimodal learning involves integrating information from multiple modalities to enhance learning accuracy and robustness \cite{7zhang2024deep}. 
Compared to unimodal approaches, multimodal methods help mitigate noise and ambiguity inherent in single-modality data \cite{8ngiam2011multimodal}. 
Common techniques include early fusion (concatenating features from different modalities) and late fusion (combining modality-specific predictions) \cite{19gadzicki2020early}. 
Multimodal learning has achieved significant success in emotion recognition, particularly when integrating physiological and behavioral signals such as EEG and eye-tracking features \cite{11zheng2018emotionmeter}.

In the context of emotion recognition, cross-modal approaches such as transformer-based architectures have gained traction due to their ability to capture long-range dependencies \cite{12tsai2019multimodal}. 
However, MLPs often outperform transformers in scenarios with limited training data or high-dimensional multimodal inputs \cite{8ngiam2011multimodal}. 
Therefore, MLP-based cross-modal method is particularly suitable for the SEED-IV \cite{11zheng2018emotionmeter} dataset, which integrates EEG and eye-tracking data.

\vspace{-10pt}

\section{Method}

\begin{figure} 
\centering 
\includegraphics[width=0.8\textwidth]{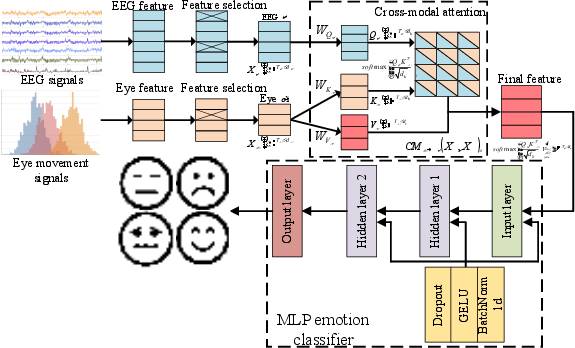}\\ 
\caption{Architecture of the proposed framework. }
\label{method_framework} 
\end{figure}

Fig. \ref{method_framework} displays the architecture of the proposed framework.
The framework comprises of data preprocessing, feature extraction, feature selection, cross-modal attention, and MLP-based emotion recognition.

\vspace{-10pt}

\vspace{-30pt}
\subsection{Data Preprocessing and Feature Extraction}

We utilize the publicly available SEED-IV dataset\footnote{https://bcmi.sjtu.edu.cn/home/seed/seed-iv.html}, which contains simultaneous EEG and eye-movement recordings from 15 users while watching emotion-eliciting videos labeled as happy, sad, fearful and neutral. 
We extract Differential Entropy (DE) features at five different frequency bands from EEG signals.
The DE of a one-dimensional signal $X$ following a Gaussian distribution $N\left( \mu ,{{\sigma }^{2}} \right)$ is determined:

\begin{equation}
h\left( X \right)=-\int_{-\infty }^{\infty }{P\left( x \right)\log \left( \text{P}\left( X \right) \right)}=\frac{1}{2}\log 2\pi e{{\sigma }^{2}}.
\end{equation}

The resulting EEG feature vector has a dimension of 310 (62 channels x 5 frequency bands).
The dataset provides 31 computational features related to eye movement, including pupil diameter, dispersion, fixation, and blink.

\vspace{-6pt}
\subsection{Feature Selection}
\vspace{-5pt}
To optimize multimodal emotion recognition using EEG and eye movement features, we employ a hybrid feature selection pipeline \cite{31hag2021enhancing} that integrates normalization, dimensionality reduction, and discriminative feature selection.
The specific steps for feature selection are as follows.
(1) The process begins with temporal normalization via Min-Max scaling applied to the raw EEG and eye-movement features.
(2) The main selection mechanism utilizes Recursive Feature Elimination (RFE) with a linear SVM.
For EEG, 200 discriminative features are selected by progressively eliminating redundant channels.  
For eye-movement data, 20 optimal features are retained based on their relevance to emotion classification.
(3) At the same time, Principal Component Analysis (PCA) is performed to reduce feature dimensionality while preserving the global feature structure.  
Specifically, 80 principal components are retained for EEG.
(4) The final feature representation is constructed by concatenating the PCA-reduced components and the RFE-selected raw features, resulting in fused feature dimensions of 280 for EEG and 51 for eye movement data.  

\vspace{-6pt}
\subsection{Cross-modal Attention}
\vspace{-6pt}

The cross-modal attention mechanism dynamically aligns complementary EEG (denoted as \( \mathbf{X}_\alpha \in \mathbb{R}^{T_\alpha \times d_\alpha} \)) and eye movement (\( \mathbf{X}_\beta \in \mathbb{R}^{T_\beta \times d_\beta} \)) sequences by computing attention weights between their temporal positions. Specifically, queries (\( \mathbf{Q}_\alpha = \mathbf{X}_\alpha \mathbf{W}_{Q_\alpha} \)), keys (\( \mathbf{K}_\beta = \mathbf{X}_\beta \mathbf{W}_{K_\beta} \)), and values (\( \mathbf{V}_\beta = \mathbf{X}_\beta \mathbf{W}_{V_\beta} \)) are derived from learnable projections, where \( \mathbf{W}_{Q_\alpha}, \mathbf{W}_{K_\beta}, \mathbf{W}_{V_\beta} \) are weight matrices. The output \( \mathbf{Y}_\alpha \), sharing the temporal length \( T_\alpha \) of EEG, is computed as:  

\begin{equation}
	{{Y}_{\alpha }}=C{{M}_{\beta \to \alpha }}\left( {{X}_{\alpha }},{{X}_{\beta }} \right)=\text{softmax} \left( \frac{{{Q}_{\alpha }}K_{\beta }^{T}}{\sqrt{{{d}_{k}}}} \right){{V}_{\beta }},
\label{eq_2}
\end{equation} 

where the \( (i,j) \)-th entry in the attention matrix quantifies the relevance of the \( j \)-th eye feature to the \( i \)-th EEG timestep. This mechanism adaptively fuses EEG and eye features into a unified representation by summarizing eye movement contributions weighted by EEG-guided attention.

\vspace{-5pt}

\subsection{MLP-based Emotion Recognition}
\vspace{-5pt}
The MLP classifier adopts a deep residual network comprising four fully connected layers to map the fused cross-modal features to corresponding emotion labels. 
Prior to entering the MLP, the fused feature vector is first projected into a 128-dimensional space. 
Then the MLP processes the fused features through hidden layers of 256, 256, and 128 units, culminating in a 4-dimensional output for emotion classification.
To ensure stable training and mitigate overfitting, batch normalization, a 30\% dropout rate, and the GELU activation function are employed throughout the layers.

\vspace{-6pt}
\section{Experiments and Results}
\vspace{-5pt}
\subsection{Experiment Settings and Baseline Methods}

The SEED-IV dataset was utilized for emotion recognition.
The dataset was partitioned into train (70\%), validation (15\%), and test (15\%) sets using stratified sampling to preserve the class distribution. 
We used the PyTorch framework for model training and evaluation, employing the Adam optimizer with a learning rate of 0.0005 and 100 iterations. 
The batch size was set to 16, and 5-fold cross validation was conducted for robust evaluation.
Model performance was evaluated using four complementary metrics: accuracy, precision, recall, and F1-score. 

To validate the effectiveness of the proposed framework in emotion recognition, two baseline methods were implemented for comparative analysis. 
The first baseline was early feature fusion with an SVM \cite{11zheng2018emotionmeter}.
The second baseline was an ablated version of our framework without feature selection, where the hybrid feature selection pipeline (PCA and RFE) was excluded.

\vspace{-12pt}

\subsection{Analysis Result of the Proposed Method}
\vspace{-18pt}
\begin{table}[htbp]
  \centering
  \caption{Recognition Performance (Mean ± Standard Deviation)}
  \resizebox{\linewidth}{!}{
    \begin{tabular}{|c|c|c|c|c|}
    \hline
    \toprule
    Methods & Accuracy & F1-score & Precision & Recall \\
    \hline
    \midrule
    Feature fusion with SVM \cite{11zheng2018emotionmeter} & 76.94±2.58 & 76.80±2.66 & 77.13±2.54 & 76.94±2.58 \\
    \hline
    \midrule
    Our framework without feature selection & 87.28±1.44 & 87.24±1.45 & 87.31±1.39 & 87.28±1.44 \\
    \hline
    \midrule
    \textbf{Our framework} & \textbf{90.62±1.06} & \textbf{90.56±1.11} & \textbf{90.70±1.10} & \textbf{90.62±1.06} \\
    \hline
    \end{tabular}%
    }
  \label{tab_result}%
\end{table}%

\vspace{-8pt}
Table \ref{tab_result} presents the average and standard deviation of accuracy, precision, recall and F1-score for three methods on the SEED-IV dataset. 
The proposed framework achieves state-of-the-art performance with 90.62\% accuracy and 90.56\% F1-score, significantly outperforming both baselines.
Compared to the early feature fusion with SVM baseline, our framework achieves improvements of 13.68\% in accuracy and 13.76\% in F1-score. 
Even without feature selection, our framework still improves accuracy by 10.34\% compared to the SVM-based baseline method, demonstrating the strong representational power of deep fusion and MLP classification. 
Further incorporating PCA and RFE feature selection leads to an additional 3.34\% gain in accuracy, underlining the effectiveness of removing noisy or redundant features. 

\vspace{-12pt}

\begin{figure}[h]
	\centering
	\begin{minipage}{0.30\linewidth}
		\vspace{3pt}
		\centerline{\includegraphics[width=\textwidth]{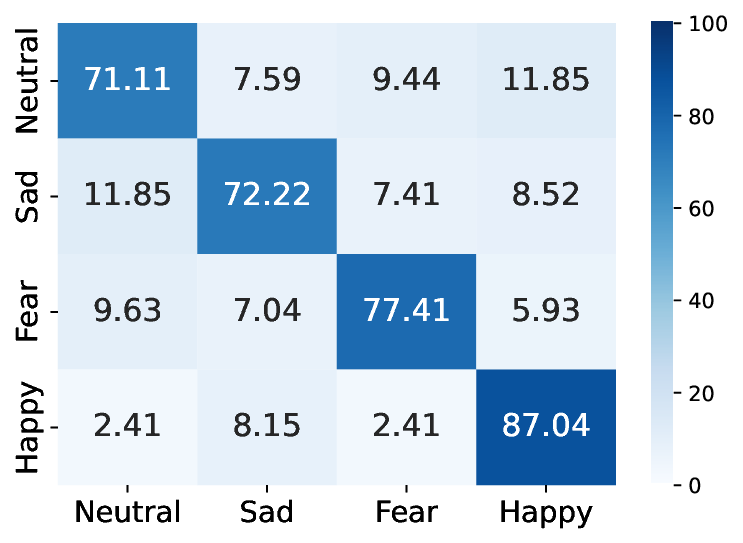}}
		\centerline{(a)}
	\end{minipage}
	\begin{minipage}{0.30\linewidth}
		\vspace{3pt}
		\centerline{\includegraphics[width=\textwidth]{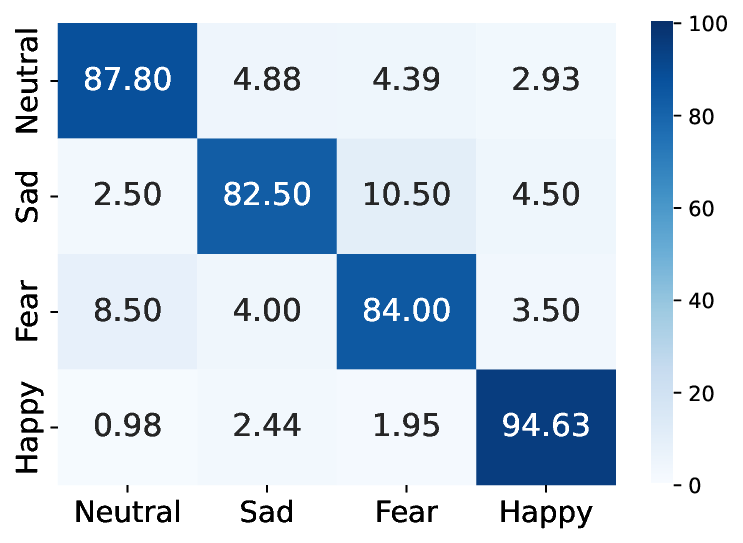}}
	 
		\centerline{(b)}
	\end{minipage}
	\begin{minipage}{0.30\linewidth}
		\vspace{3pt}
		\centerline{\includegraphics[width=\textwidth]{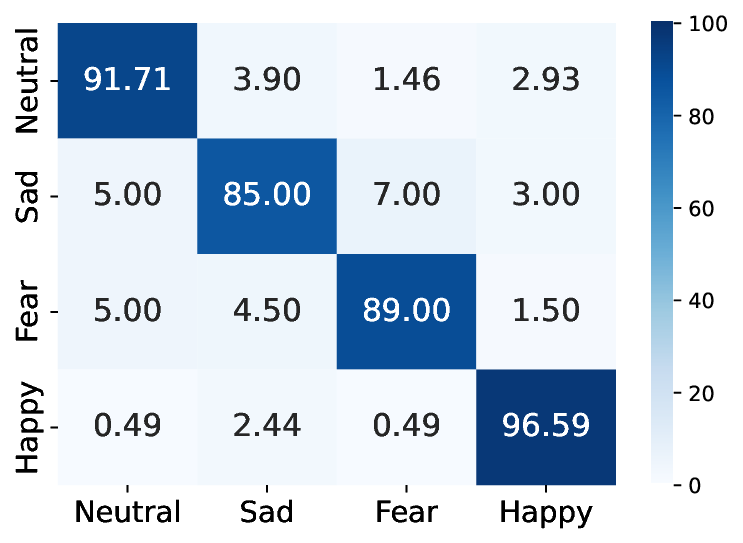}}
	 
		\centerline{(c)}
	\end{minipage}
 
	\caption{Confusion matrices for different methods. (a) Feature fusion with SVM \cite{11zheng2018emotionmeter}. (b) Our framework without feature selection. (c) Our framework.}
	\label{fig4}
\end{figure}
 
\vspace{-6pt}
Fig. \ref{fig4} shows the per-class confusion matrices for the different methods. 
All three methods achieve the highest accuracy rate in identifying the "happy" emotion among the four categories. 
Notably, our framework achieves recognition rates exceeding 85\% for all four emotions, with 91.71\% for "neutral" and an impressive 96.59\% for "happy".
Moreover, our framework improves the accuracy of neutral from 71.11\% to 91.71\%, highlighting the significance of feature selection, cross-modal attention and MLP classification.

\vspace{-10pt}
\section{Conclusion}
\vspace{-8pt}
In this paper, we proposed a cross-modal emotion recognition framework to address noise in raw multimodal data and limited temporal interaction modeling. 
Experiments demonstrated its superiority, achieving state-of-the-art performance with 90.62\% accuracy and 90.56\% F1-score. 
In future work, we will focus on advanced feature selection strategies and temporal-aware modeling to further improve accuracy and generalizability in real-world applications.

\vspace{6pt}
\noindent
\textbf{Acknowledgment}
This work was supported by the China National College Student Innovation and Entrepreneurship Training Program (Grant No. 202410424066).

\vspace{-8pt}
%
%
%
%
\begingroup
\setlength{\baselineskip}{0.2\baselineskip} 
\bibliographystyle{splncs04}
\bibliography{rfe}
\endgroup
\end{document}